\documentclass[twocolumn,english, twocolumns,superscriptaddress]{revtex4-1}
\usepackage{amsmath, amsthm, amssymb,graphicx, bbm}
\usepackage[normalem]{ulem}

\usepackage{umoline}
\usepackage[usenames,svgnames]{xcolor}

\usepackage{hyperref}
\hypersetup{
     colorlinks=true,           
     linkcolor=Crimson,          
     citecolor=blue,            
     filecolor=blue,            
     urlcolor=cyan,             
}

\DeclareMathOperator{\Tr}{Tr}
\newcommand{\ket}[1]{|#1\rangle}
\newcommand{\bra}[1]{\langle#1|}
\newcommand{\av}[1]{\langle#1\rangle}
\newcommand{\braket}[2]{\langle#1|#2\rangle}

\newcommand{\broket}[3]{\bra{#1}#2\ket{#3}}
\newcommand{\norm}[1]{\left\Vert #1 \right\Vert}
\newcommand{\orderof}[1]{\mathcal{O}(#1)} 
\newcommand{\EqDef}{\stackrel{\mathrm{def}}{=}}

\newcommand{\Eq}[1]{Eq.~(\ref{#1})}
\newcommand{\Fig}[1]{Fig.~\ref{#1}}
\newcommand{\Ref}[1]{Ref.~\cite{#1}}

\newcommand{\App}[1]{Appendix~\ref{#1}}

\begin{document}

\title{Learning a Local Hamiltonian from Local Measurements}

\author{Eyal Bairey}
\affiliation{Physics Department, Technion, 3200003, Haifa, Israel}

\author{Itai Arad}
\affiliation{Physics Department, Technion, 3200003, Haifa, Israel}

\author{Netanel H. Lindner}
\affiliation{Physics Department, Technion, 3200003, Haifa, Israel}

\begin{abstract}
Recovering an unknown Hamiltonian from measurements is an increasingly important task for certification of noisy quantum devices and simulators. Recent works have succeeded in recovering the Hamiltonian of
an isolated quantum system with local interactions from long-ranged correlators of a single eigenstate. Here, we show that such Hamiltonians can be recovered from local observables alone, using computational and measurement resources scaling linearly with the system size. In fact, to recover the Hamiltonian acting on each finite spatial domain, only observables within that domain are required. The observables can be measured in a Gibbs state as well as a single eigenstate; furthermore, they can be measured in a state evolved by the Hamiltonian for a long time, allowing to recover a large family of time-dependent Hamiltonians. We derive an estimate for the statistical recovery error due to approximation of expectation values using a finite number of samples, which agrees well with numerical simulations.
\end{abstract}
\maketitle

\paragraph*{Introduction.} 
Contemporary condensed matter physics has witnessed great advancements in tools developed to obtain the state of a system given its Hamiltonian.  As quantum devices are being rapidly developed, the converse task of recovering the Hamiltonian of a many-body system from measured
observables is becoming increasingly important. In particular, it is a necessary step for certifying quantum simulators and devices
containing many qubits. As these expand beyond the power
of classical devices \cite{Preskill2018}, there is a growing need to
certify them using only a polynomial amount of classical
computational resources as well as quantum measurements. 

Various methods have been suggested for recovering a Hamiltonian
based on its dynamics~\cite{Burgarth2009, DiFranco2009, Zhang2014,
DeClercq2016, Sone2017, Sone2017a, Wang2018} or Gibbs state ~\cite{Rudinger2015, Kieferova2017, Kappen2018}. The system-size scaling of the recovery efficiency can be improved using a trusted quantum simulator~\cite{Granade2012, Wiebe2014,
Wiebe2014a, Wiebe2015, Wang2017}, manipulations of the investigated
system ~\cite{Wang2015}, or accurate measurements of short-time
dynamics \cite{Shabani2011a, DaSilva2011}.

Here, we suggest a framework for recovering a generic local
Hamiltonian using only polynomial time and measurements. Inspired by the
recently introduced method for recovering a local Hamiltonian from measurements on a
single eigenstate ~\cite{Qi2017,Chertkov2018, Greiter2018}, our
framework offers four main contributions. First, we generalize to mixed states such as Gibbs states $\rho=\frac{1}{Z}e^{-\beta H}$, treating any 
state which commutes with the Hamiltonian at the same footing as an
eigenstate. Second, our method can be applied to dynamics of arbitrary low-energy initial states time-evolved by the Hamiltonian. Third, it allows to recover time-dependent Hamiltonians if the functional form of their
time-dependence is known. Finally, in the case of short-range interactions, we can infer the Hamiltonian of a
local patch $L$ based only on local measurements inside $L$. This implies that a short-ranged Hamiltonian on a large system can be obtained with a number of measurements and computation time \emph{linear} in system size.

\paragraph*{Problem setting.}

\begin{figure}[h!]
  \includegraphics[width=8.6cm]{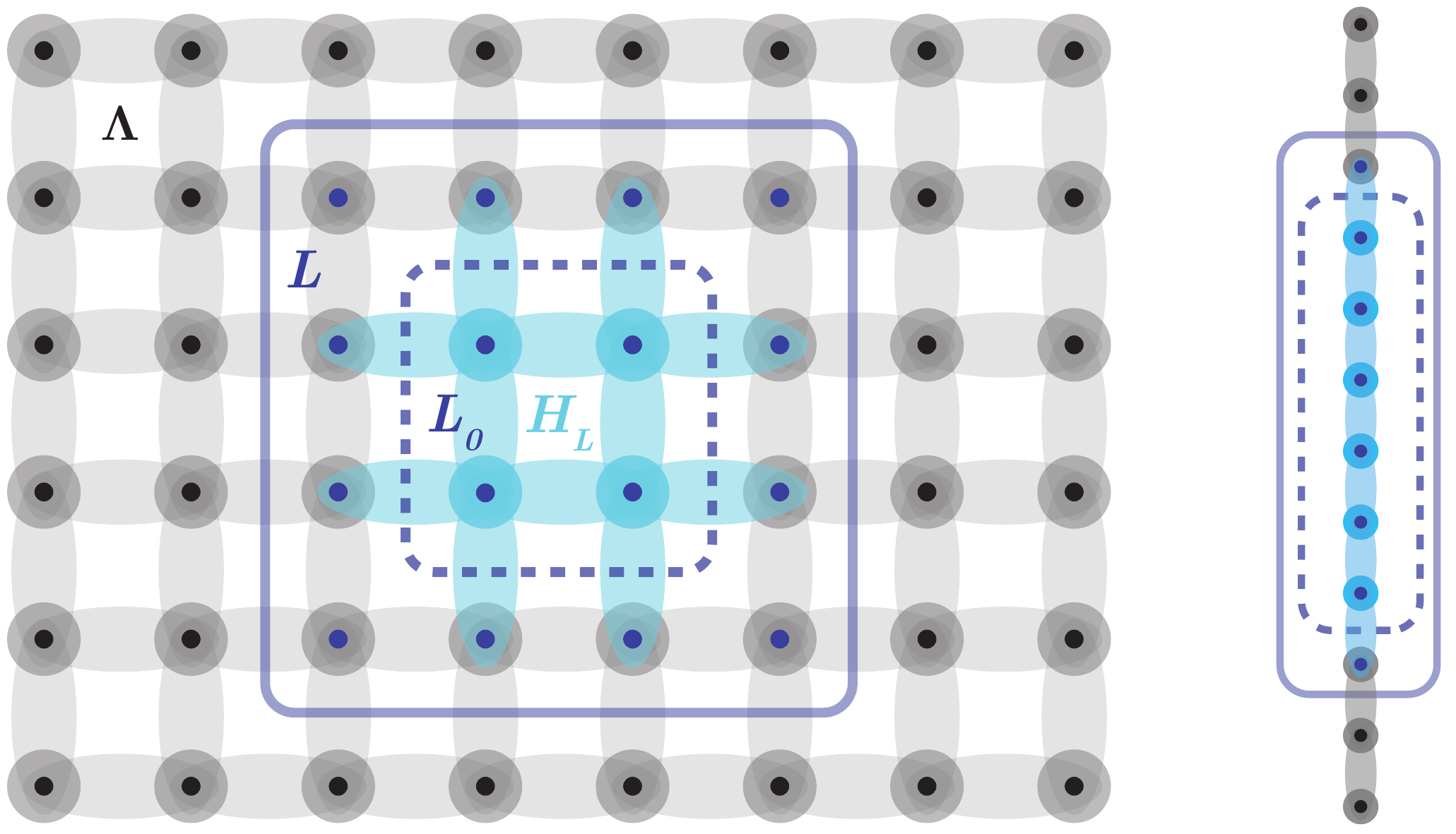} 
  \protect
  \caption{ Recovery from local measurements. Left: our method
    recovers $H_L$ (light blue), using only measurements of
    observables residing in $L$ (solid blue line), a sub-region of the whole
    system $\Lambda$. The interior $L_0\subseteq L$ (dashed blue line) consists of sites interacting only within $L$.
    Right: our simulations are performed on $|\Lambda|=12$ chains,
    recovering $H_L$ on the 8 middle spins. } 
 \label{fig1}
\end{figure}

We wish to recover the Hamiltonian acting on a
region $L$ by measuring observables only in $L$.
We would first like to make these notions precise.

We consider a Hamiltonian $H$ on a finite lattice $\Lambda$
in $d$ dimensions:
\begin{equation}
  H = \sum_i h_i.
  \label{eq:H_local}
\end{equation}
We assume that $H$ is $k$-local, such that each $h_i$ acts
non-trivially on no more than $k$ spatially contiguous sites (i.e., contained
within a ball of diameter $k$). We focus on a specific subset of
sites $L\subseteq \Lambda$. We define its interior $L_0 \subseteq L$ as the sites that are not conected by $H$ to sites outside $L$ (Fig. \ref{fig1}). We denote by $H_L$ the
subset of $h_i$ terms in $H$ that act non-trivially on $L_0$. 

We call any state $\rho$ that is stationary under $H$ a
\textit{steady state} (taking $\hbar=1$): 
\begin{equation}
  i \partial_t \rho = \left[H ,\rho \right] = 0,
\label{eq:eq1}
\end{equation}
In particular, $\rho$ can be any eigenstate as well as a Gibbs
state. Our goal is to recover $H_L$ from a steady state of $H$, based only
on measurements in $L$.

\paragraph*{Algorithm.} 

To recover $H_L$, we identify a set of local
constraints on $H_L$ obeyed by any steady state $\rho$ of $H$. Since
$\rho$ is stationary under $H$, so is the expectation value
$\left\langle A \right\rangle \EqDef \Tr \left( \rho A \right)$ of
any operator $A$ in the state $\rho$, so that $\partial_t \av{A}  = -\av{i [A,H]}= 0$. If $A$ is supported only on $L_0$,
this constraint becomes 
\begin{equation}
\av{i[A,H_L]}=0,
\label{constraint}
\end{equation} 
since $A$ trivially commutes with $H-H_L$. 

The $k$-local operators acting on $L_0$ form a linear space. We choose a basis $\left\lbrace S_m \right\rbrace_{m=1}^M$ for this space of operators, where $M$ is its dimension. When we expand $H_L$ in this basis,
\begin{equation}
  H_L = \sum_{m=1}^M c_m S_m,
\end{equation}
the constraint \eqref{constraint} becomes a linear
homogeneous constraint on the vector $\vec{c} = (c_1, c_2, \ldots,
c_M)$:
\begin{equation}
  \sum_{m=1}^M c_m \av{i[A, S_m]} = 0 .
\end{equation}

Using a set of operators $\left\lbrace A_n \right\rbrace_{n=1}^N$,
each supported on $L_0$, we obtain a set of $N$ linear constraints:
\begin{equation}
  \forall n: \sum_{m=1}^M c_m 
    \left\langle i \left[A_n, S_m\right] \right\rangle  = 0,
\end{equation}
which is equivalent to the $N \times M$ real linear equation
\begin{align}
  K \vec{c} &= 0, &  
  K_{n,m} &\EqDef \av{i[A_n,S_m]} .
\label{eq:K}
\end{align}
The number $M$ of basis elements $S_m$ that span $H_L$ is linear in the subsystem's volume $|L|$. In contrast, the maximal number of constraints 
scales like the number of linearly independent observables $A_n$ in
$L_0$, which grows exponentially with $|L_0|$. Thus, for a sufficiently
large but constant region $L$ (depending on $k$ but not on $|\Lambda|$), we can always have more equations
than unknowns, i.e., $N>M$. As argued in
\Ref{Qi2017}, we expect these equations to be generally independent,
thereby providing a unique solution $\vec{c}$ up to an overall scale.

Given a region $L$ whose Hamiltonian we wish to learn, our method is
therefore as follows: 
\begin{enumerate}
  \item Identify a set of terms $\{S_m\}_{m=1}^M$ 
    spanning the space of possible $H_L$'s.
    
  \item Construct a constraint matrix $K_{N \times M}$ by 
    measuring $\av{i[ A_n, S_m]}$ with respect to a set of
    constraints $\{A_n\}_{n=1}^N$ supported on $L_0$.
    
  \item Estimate $H_L \propto \sum_{m=1}^Mc_m S_m$, with $\vec{c}$
    the lowest right-singular vector of $K$.
\end{enumerate}

The lowest right-singular vector of $K$ is the numerical solution to
\Eq{eq:K}, the vector that minimizes $\norm{Kc}$. Namely, it is the
ground-state of the \textit{correlation matrix},
\begin{equation}
  \mathcal{M}=K^T K.
  \label{Eq:M}
\end{equation}

\paragraph*{Extension to a dynamical setting.} So far, we have described how to recover a time-independent $H$ from measurements of its steady state. However, many experimental settings do not have access to an exact steady state of $H$. Instead, we now describe how to obtain an approximate steady-state from an arbitrary initial state by evolving it with $H$ for long times. 

In the dynamical approach, we repeatedly initialize our system in some
state $\rho\left(0\right)$. We let it evolve for a random time distributed uniformly in $0 \leq t' \leq t$, before measuring an operator $A$. The average
outcome of these measurements is given by $\Tr\left(
\rho_{avg} A \right) $, where 
$\rho_{avg}=\frac{1}{t}\intop_{t'=0}^t\rho (t') dt'$. For a time-independent $H$, this time-averaged density matrix 
approaches a steady state in trace norm, since by integrating \eqref{eq:eq1}, we obtain:
\begin{equation}
  \norm{\left[\rho_{avg},H \right]}_1 = 
  \frac{1}{t} \norm{  \rho (t) - \rho (0) }_1 \leq 
  \frac{2}{t}.
  \label{thermalization}
\end{equation}
This allows to recover a time-independent $H$ from a constraint matrix $K$ of time-averaged observables.

The dynamical approach can be extended to time-dependent
Hamiltonians of the form:
\begin{equation}
  \hat{H}(t) = \hat{H}^{(0)} + \hat{V}f(t),
\end{equation}
where $f(t)$ is a known function. Similarly to \Eq{thermalization},
now the time-averaged commutator $\frac{1}{t}\intop_0^t[\rho(t'), \hat{H}(t')]dt'$
must decay with time. Therefore, we estimate the coefficients of $\hat{H}^{(0)}$, $\hat{V}$ as the lowest singular vector of an extended constraint matrix $K_{N \times
2M}$ composed of time-averaged as well as time-modulated
measurements (see \App{app_td}):
\begin{equation}
\begin{aligned}
  \forall m \leq M: \ K_{n,m} 
    &= \frac{1}{t}\intop_{0}^t\av{[A_n,S_m]} dt' \\
  K_{n,m+M} &= \frac{1}{t}\intop_{0}^t
    \av{[A_n,S_m]} f(t') dt'
\label{floquet}
\end{aligned}
\end{equation}

\paragraph{Sample complexity.} 
The complexity of our method depends on the number of observables we
need to measure and on the accuracy to which we need to measure each of
them. Experimentally, each observable $\av{i[A_n, S_m]}$ can only be measured to finite
accuracy due to statistical uncertainty in estimating it using a finite number of samples $n_s$. We quantify the
resulting error in the reconstruction process by the $l_2$ distance
between the normalized recovered and true coefficient vectors
\footnote{Equivalently, $\Delta = 2| \sin{\frac{\theta}{2}} |$, where $\theta$ is the angle between the two vectors
\cite{Qi2017}}, 
\begin{equation}
  \Delta = \norm{\hat{c}_{true} - \hat{c}_{recovered}}_2,
\label{Eq:l2}
\end{equation}
where $\hat{c} \EqDef \frac{\vec{c}}{\norm{\vec{c}}}$.

Following {\Ref{Qi2017}}, we analyze the
 reconstruction error using a simple perturbation theory on
the correlation matrix {$\mathcal{M}$}. We model the error in each entry $K_{n,m}$ obtained by $n_s$ samples as an 
independent Gaussian with zero mean and standard deviation $\epsilon
\approx n_s^{-1/2}$. To lowest order in $\epsilon$, we 
estimate the expected error: 
\begin{equation}
  \mathbb{E}(\Delta) \approx \epsilon \sqrt{\sum_{i>0} 
    \frac{1}{\lambda_{i}}} \EqDef \Delta^{est},
\label{error_estimate}
\end{equation}
where $\lambda_{i}$ are the eigenvalues of $\mathcal{M}$ (see \App{app_err}).

To open a gap in $\mathcal{M}$ between $\lambda_0$ and $\lambda_1$ and
recover a unique Hamiltonian, at least as many constraints $N$ as
unknowns $M$ are required. This means measuring $\orderof{|L|}$ operators $i[A_n, S_m]$, since each constraint $A_n$ commutes with all but a constant number of
candidate Hamiltonian terms $S_m$. Moreover,
the Hamiltonian can be reconstructed in linear time
in $|L|$ and a linear
number of measurements by breaking down $L$ into smaller
sub-regions and reconstructing the Hamiltonian on each of them
separately. For translationally invariant Hamilto-
nians  a  single  sub-region  is  sufficient,  with  only  a  constant number of operators to be measured.

Minimizing the support of the measured operators is advantageous for some experimental settings, in which correlations involving multiple sites are hard to measure. Suppose, for example, we wish to recover a generic 2-local $H$. To obtain more equations than unknowns, we need constraints $A_n$ that act on at least 2 sites. This corresponds to 3-local measurements $i[A_n, S_m]$. Luckily, measurements of only 2-local observables can suffice if a few different steady states are available. These may be Gibbs states at different temperatures, or time-averaged evolutions of different initial conditions. In this setting, each steady state can provide an independent set of constraints. For a 2-local Hamiltonian in one dimension, single-site $A_n$ operators and 5 different steady states can provide sufficient constraints to open a gap in $\mathcal{M}$. More generally, access to multiple steady states allows to recover a $k$-local $H$ using only $k$-local measurements.

\paragraph{Numerical simulations.} 

\begin{figure}[t]
\includegraphics[width=8.6cm]{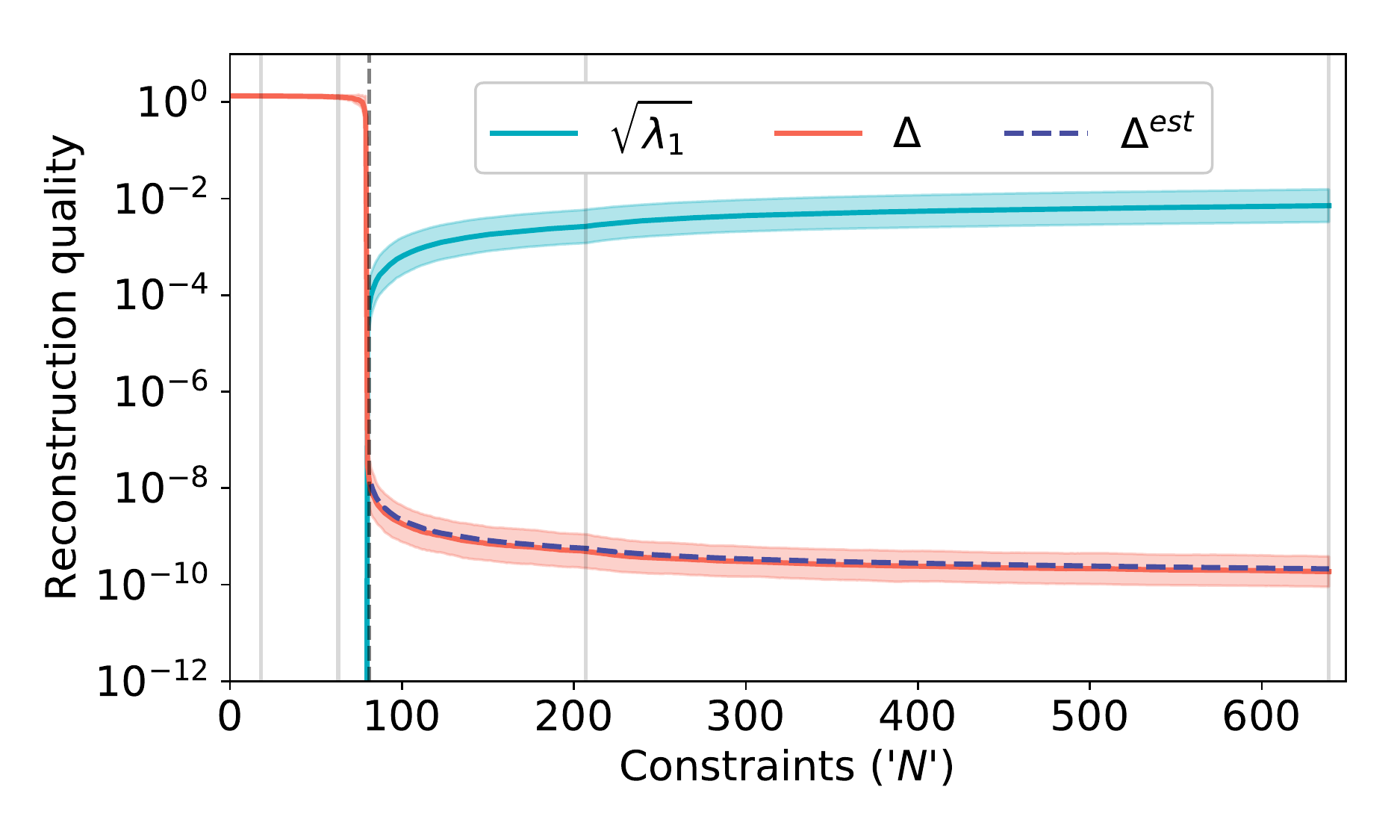} \protect 
  \caption{Quality of
  Hamiltonian reconstruction as a function of the number of measured
  constraints $N$. We generated ground-states of random spin chains
  (\Eq{Eq:H_random}), and measured local observables $i[A_n, S_m]$ on the 8 middle spins $L$ to recover $H_L$. When the number $N$ of
  constraints $A_n$ exceeded the number $M$ of possible Hamiltonian
  terms $S_m$ (dashed vertical line), $\mathcal{M}$ opened a gap (light blue; $\lambda_0=0$ here). The reconstruction error [red, see Eq.
  \eqref{Eq:l2}] was solely due to the addition
  of a small Gaussian noise with standard deviation $\epsilon=10^{-12}$ to each measurement. THe error closely followed an estimate obtained from the spectrum of $K$ [dashed purple, see Eq. \eqref{error_estimate}].
  We used all $k$-local constraints $A_n$ up to $k=4$ in an increasing order of support size $k$. The solid vertical lines denote the transition to
  $k=2,3,4$ respectively, and within each $k$ we chose the constraints in random order. Results were averaged
  over $200$ random Hamiltonians; the means and standard deviations were calculated after taking the log.} \label{fig2}
\end{figure}

To demonstrate the performance of our method, we numerically simulated random one-dimensional spin $\frac{1}{2}$
chains. We considered Hamiltonians consisting of all possible 2-local
terms, acting on single spins and nearest neighbors:
\begin{equation}
  H = \sum_{l=1}^{|\Lambda|}  
    \sum_{\alpha=1}^{3} c_{l\alpha} \sigma^{\alpha}_l + 
    \sum_{l=1}^{|\Lambda|-1} 
    \sum_{\alpha=1}^{3} \sum_{\beta=1}^{3} 
    c_{l \alpha \beta} \sigma^{\alpha}_l \sigma^{\beta}_{l+1} .
\label{Eq:H_random}
\end{equation}

In each simulation, we generated a random 2-local Hamiltonian $H$
(Eq. \eqref{Eq:H_random}) on $|\Lambda|=12$ sites by sampling the
vector of all coefficients $\vec{c}$ from a Gaussian distribution with zero mean and unit standard deviation, setting the energy scale for what follows. We numerically calculated the ground-state of $H$, and then recovered $H_L$ from the ground-state in steps. In each step we added
one row to the constraint matrix $K$ by choosing a constraint
operator $A_n$ and estimating $\{\av{i [A_n,S_m]}\}_{m=1}^M$. Here, $A_n$ is an operator supported on the 6 middle sites $L_0$, and $\left\lbrace S_m \right\rbrace_{m=1}^M$ is the subset of terms in \Eq{Eq:H_random} acting on $L_0$. To measure the robustness of the reconstruction, we added to the
constraint matrix $K$ a noise matrix of independent Gaussian entries with
zero mean and standard deviation $\epsilon = 10^{-12}$.

As expected, once sufficiently many constraints had been measured,
our procedure recovered the Hamiltonian to high accuracy (Fig.
\ref{fig2}). As soon as $N=M-1$, the correlation matrix
$\mathcal{M}$ opened a gap, allowing to recover the coefficient
vector $\hat{c}$ given by the ground-state of $\mathcal{M}$. As more constraints were added, the gap
gradually grew. The reconstruction error decreased correspondingly, showing excellent agreement with our estimate 
\eqref{error_estimate}. We also ran simulations on random $XY$ chains to reach larger system sizes ($|\Lambda=100|$). The gap of the correlation matrix seemed insensitive to the size of the sub-system for the range we examined $7 \leq |L| \leq 13$ (\Fig{figS1} in \App{app_XY}). 

\paragraph*{Reconstruction from Gibbs states.} 
Next, we reconstructed $H_L$ for random spin chains from measurements of their Gibbs states. We sampled
200 random Hamiltonians \eqref{Eq:H_random} on $|\Lambda|=12$ sites
and generated Gibbs states $\frac{1}{Z}e^{-\beta H}$ for varying $\beta\in
\left[0.01, 1 \right]$. We then measured a fixed number of
observables, corresponding to all 4-local constraints $A_n$ supported on the 6 middle spins $L_0$. We added a small noise ($\epsilon=10^{-12}$) to each measurement. 

\begin{figure}[t]
\includegraphics[width=8.6cm]{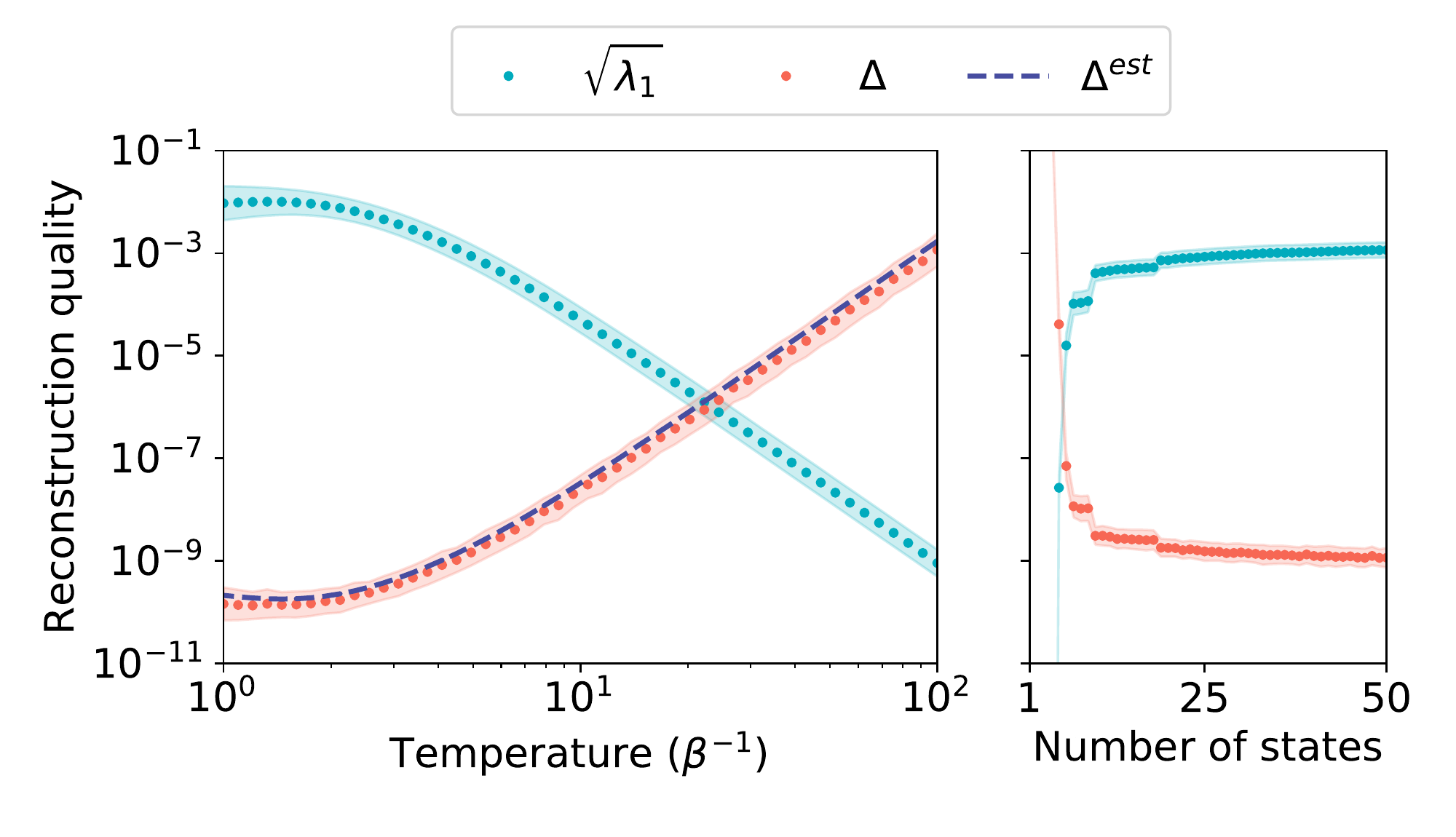} 
\protect\caption{Quality of
Hamiltonian reconstruction from Gibbs states $\rho=\frac{1}{Z}e^{-\beta H}$. We reconstructed $H_L$ on the $L=8$ middle spins of random spin chains \eqref{Eq:H_random} of length $|\Lambda|=12$ (see Fig.
\ref{fig1}). Left: as
a function of temperature $T=\beta^{-1}$, using 4-local constraints $A_n$. The gap of the correlation matrix $\mathcal{M}$
decreased with temperature (light blue). Correspondingly, the
reconstruction error (red) due to a small measurement uncertainty ($\epsilon=10^{-12}$) increased according to the estimate \eqref{error_estimate} obtained from the
spectrum of $K$ (dashed purple). Right: reconstruction with 2-local measurements only, using single-site constraints $A_n$ and multiple Gibbs states of different temperatures. We generated Gibbs states at temperatures in the range $\beta^{-1}=[10^0, 10^2]$, chosen with uniform spacings (in log space) which decreased with the number of states. A few different states sufficed; additional states improved the reconstruction quality. Results were averaged over 200 randomizations.} \label{fig3}
\end{figure}

Our results show that the reconstruction error increases with temperature
(Fig.~\ref{fig3}, left). As the
system approaches a fully mixed state, the commutator $[H,\rho]$ 
approaches zero for \emph{every} $H$, which implies that many
different $H$ are becoming compatible with $\rho$.
Correspondingly, the elements of the constraint matrix $K$ shrink,
and so does its gap. At low temperatures, the reconstruction quality was similar to that of ground-states. By combining measurements performed at different temperatures, we were able to recover $H$ using only 2-local measurements (Fig.~\ref{fig3}, right).

\paragraph*{Reconstruction from dynamics.} 
To demonstrate Hamiltonian recovery from the dynamics of an initial state, we simulated a quench protocol. We generated two random Hamiltonians $\hat{H}^{(0)}$, $\hat{H}^{(1)}$ on $|\Lambda|=12$ sites from the ensemble
\eqref{Eq:H_random}. We initialized our system in the ground-state of
$\hat{H}^{(0)} + \hat{H}^{(1)}$, and evolved it by $\hat{H}^{(0)}$
alone. This initialization yielded states whose energy with respect to the final Hamiltonian was not too high. We then attempted at different
times $t$ to recover $H_L^{(0)}$ on the 8 middle spins using 4-local constraints $A_n$. We
did this by constructing a constraint matrix $K_t$ from time-averaged values of $\av{i [A_n, S_m ]}$, sampled at equally spaced
intervals $dt=0.05$ up to time $t$. 

\begin{figure}[t]
\includegraphics[width=8.6cm]{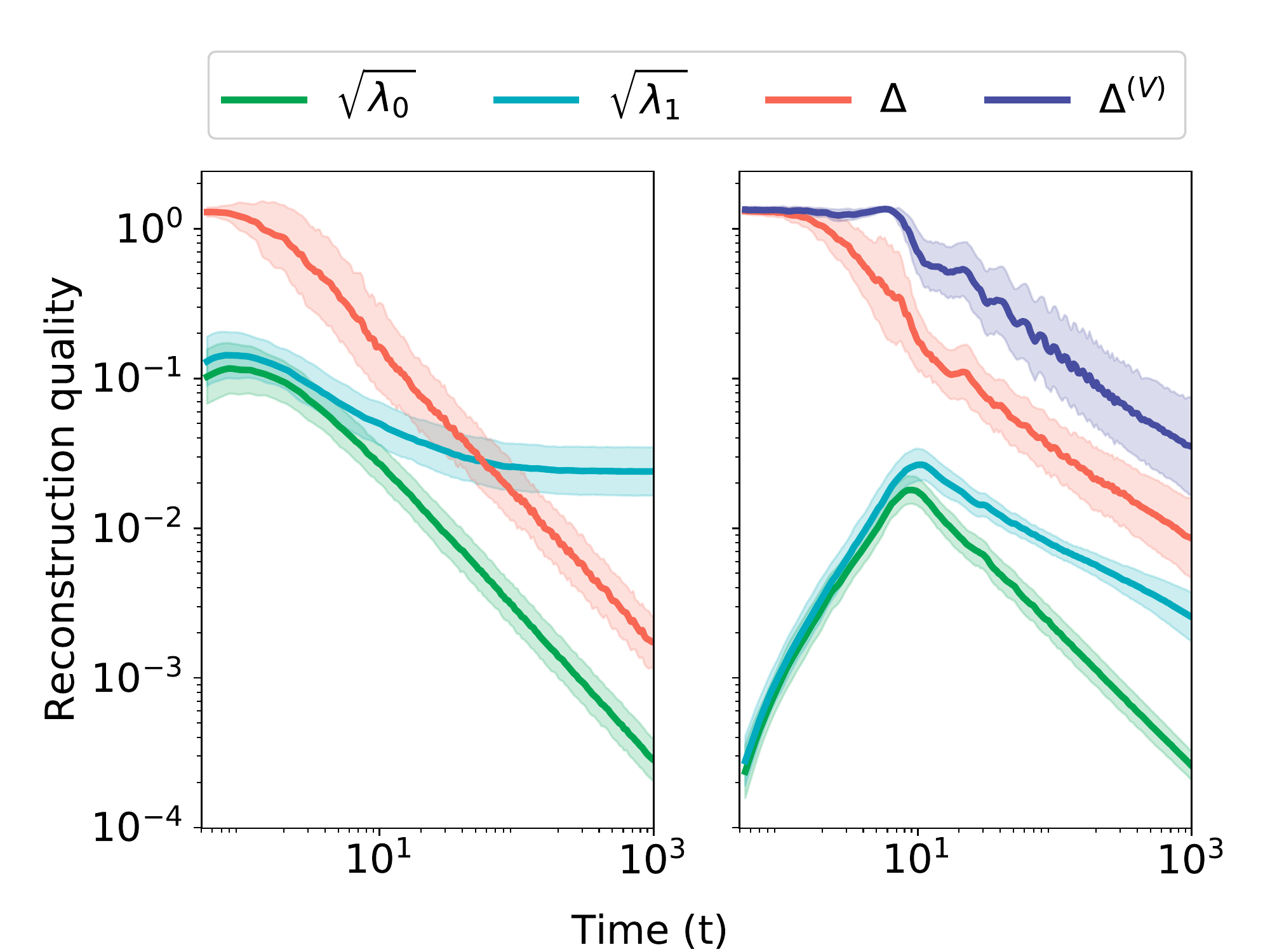} \protect\caption{Reconstruction from dynamics, as a function of time, of the final Hamiltonian following a quench at $t=0$.  Left:
reconstruction of a time-independent Hamiltonian from $\rho_{avg}(t) = \frac{1}{t} \intop_0^t \rho(t') dt'$. While the first excited eigenvalue $\lambda_1$ of the correlation matrix $\mathcal{M}$ saturated (light blue), its lowest eigenvalue
$\lambda_0$ decayed with time (green), leading to a decrease in the reconstruction error $\Delta$ (red). Right: reconstruction of a
time-oscillating Hamiltonian 
$\hat{H}(t) = \hat{H}^{(0)} + J\hat{V} \cos{\omega t}$. Here $\lambda_1$
decreased with time due to heating, leading to a larger reconstruction error $\Delta$ compared to the time-independent case (red for $\hat{H}^{(0)}$, purple for $\hat{V}$). Results
were averaged over 50 randomizations. } \label{fig4}
\end{figure}

After a transient period, the first excited eigenvalue $\lambda_1$ of the correlation matrix $\mathcal{M}$ saturated (Fig. \ref{fig4}). The lowest eigenvalue $\lambda_0$ continued to decay, opening a gap which widened with time. This decay fits to the power law $\sqrt{\lambda_0} \propto \frac{1}{t}$, reflecting the
expected decay rate of the commutator with the true Hamiltonian from Eq. \eqref{thermalization}. Here the finite value of $\lambda_0$ played the role of noise, leading to reconstrcution error. As $\lambda_0$ decreased, the Hamiltonian was reconstructed to better and better
accuracy. 

\paragraph*{Recovery of time-dependent Hamiltonians.} 

We repeated the quench experiments with a final Hamiltonian which
depends on time, focusing on a periodic drive with a single
frequency: $f(t)=\cos{\omega t}$. We initialized our system in the
ground-state of $\hat{H}^{(0)} + \hat{H}^{(1)}$ and evolved it in
time with 
\begin{equation}
  \hat{H}(t) = \hat{H}^{(0)} + J \hat{V} \cos{\omega t}, 
\end{equation}
taking $J=0.5$ and $\omega=0.05$. We sampled all three terms
$\hat{H}^{(0)}, \hat{H}^{(1)}$ and $\hat{V}$ identically using the form given by Eq.
\eqref{Eq:H_random}. We then constructed at different times an extended
constraint matrix $K_{N\times 2M}$ from time-averaged as well as
time-modulated observables [see Eq. \eqref{floquet}].

As in the time-independent case, $\lambda_0$ decayed with time (Fig. \ref{fig4}, right). However, $\lambda_1$ decayed too, indicating a small or vanishing gap $\lambda_1-\lambda_0$ for long times, corresponding to high temperatures (see Fig. \ref{fig3}). Recovery in this case is therefore possible when the system does not heat too quickly, i.e. when $\lambda_1$ decays slower
than $\lambda_0$, which depends on the driving amplitude $J$ and
frequency $\omega$ (see \App{app_drive}).

\paragraph*{Discussion.} 
We suggest a framework for inferring local Hamiltonians. Our
framework generalizes the recently-introduced correlation matrix
formalism \cite{Qi2017, Chertkov2018, Greiter2018}, applying to
Gibbs states and dynamics as well as eigenstates. Importantly, it allows to recover short-ranged
Hamiltonians using measurements as well as computational resources scaling linearly with system size. 

We point out that even when the available measurements do not provide sufficient constraints to open a gap in $\mathcal{M}$ and yield a unique Hamiltonian $H$, our method recovers a linear subspace containing $H$. This can be combined with additional knowledge, e.g. to verify the accuracy of a prior guess for $H$ or to improve such a guess.

Most of our formalism applies equally well to long-ranged
Hamiltonians, in which interactions can involve
any arbitrary set of $k$ spins. Our algorithm must then be applied
to the whole system $\Lambda$ at once rather than locally. Still, the number of
possible Hamiltonian terms scales polynomially with system size, as
$|\Lambda|^{k}$. 

Note that when we enforce stationarity of all possible observables $A_n$ on the full system $L=\Lambda$, our correlation matrix  takes the appealing
form $\mathcal{M}^{\Lambda}_{ij} = \Tr\left( 
    \left[ \rho, S_i \right]^\dagger \left[ \rho, S_j 
      \right] \right)$, coinciding with the correlation matrix defined in \Ref{Qi2017} (up to a scalar; see \App{app_M}).  If we suffice with the full set of observables $A_n$ on the interior $L_0$ of a subsystem, \Eq{eq:K} is equivalent to the operator identity $\Tr_{\partial L}[\rho_L, H_L] = 0$. Here, $\rho_L$ is the reduced density matrix on $L$ and $\Tr_{\partial L}$ is a partial trace on the boundary spins $\partial L \EqDef L - L_0$ included in $L$ but not in its interior (see \Ref{Anshu2016}). We also note that adding constraints and Hamiltonian terms acting on $\partial L$ converts our algorithm to a method for finding the entanglement Hamiltonian on $L$ (similar to \cite{Zhu2018}).

\begin{acknowledgments}
  We thank Miklos Santha and Anupam Prakash for illuminating
  discussions, and Renan Gross for critical comments on the
  manuscript. E. B. and N. L. acknowledge financial support from the
  European Research Council (ERC) under the European Union Horizon
  2020 Research and Innovation Programme (Grant Agreement No.
  639172). I.~A. acknowledges the support of the Israel Science Foundation (ISF) under the Individual Research Grant 1778/17. N. L. acknowledges support from the People Programme
  (Marie Curie Actions) of the European Union\textquoteright s
  Seventh Framework Programme (No. FP7/2007\textendash 2013) under
  REA Grant Agreement No. 631696 and from the Israeli Center of
  Research Excellence (I-CORE) \textquotedblleft Circle of
  Light.\textquotedblright{}.
\end{acknowledgments}

\nocite{Peschel2003, Cheong2004a,
Peschel2009, Sachdev2011}

\bibliographystyle{apsrev4-1}

\bibliography{library}

\appendix

\renewcommand{\theequation}{S\arabic{equation}}
\renewcommand{\thefigure}{S\arabic{figure}}
\setcounter{equation}{0}
\setcounter{figure}{0}

\section{Recovering time-dependent Hamiltonians}
\label{app_td}

Suppose we wish to recover a time-dependent Hamiltonian of the form:
\begin{equation}
  \hat{H}(t) = \hat{H}^{(0)} + \hat{V}f(t),
\end{equation}
where $f(t)$ is a known function and $\hat{H}, \hat{V}$ are
the operators we wish to learn. For any operator $A$, Schrodinger's
equation now reads: 
\begin{equation}
  i\partial_t\av{A}  =\av{[A,\hat{H}^{(0)}]} +
  \av{[A, f(t) \hat{V}]}.
\end{equation}
Integrating the above equation and expanding in
local operators: $\hat{H}^{(0)} = \sum c^{(0)}_m h_m$, $\hat{V}=\sum
c^{(v)}_m h_m$, we obtain:
\begin{align}
\nonumber
  \left| \sum_j \frac{c_j^{(0)}}{t} 
    \intop_0^t \av{[A, h_m]}dt'
    + \sum_j \frac{c_j^{(v)}}{t} \intop_0^t 
      \av{[A, h_m]}f(t')dt'\right|\\ 
   \leq \frac{2 \norm{A}}{t}.
  \label{floquet2}
\end{align}
Minimizing the LHS of \eqref{floquet2} with respect to a set of
operators $\{ A_n \}_{n=1}^N$ amounts to finding the
lowest right-singular vector of the extended constraint matrix $K_{N\times 2M}$, defined as
\begin{align*}
  \forall m \leq M: \ K_{n,m} 
    &= \frac{1}{t}\intop_0^t \av{[A_n,h_m]} dt' \\
  K_{n,m+M} &= \frac{1}{t}\intop_0^t \av{[A_n,h_m]} f(t') dt' .
\end{align*}


\section{Error estimation}
\label{app_err}

Experimentally, each element of the constraint matrix
$K_{n,m}=\av{[A_n,S_m]}$ can only be estimated using a finite number
of samples $n_s$. Therefore, the measured empirical constraint matrix
$\hat{K}$ deviates from the true one $K$ by a noise matrix. We would
like to estimate the error in the recovered Hamiltonian due to this
noise. 

We study the effect of the noise by treating it as a perturbation.
We assume that the correct $K$ has a one-dimensional kernel; namely,
we were given a state $\rho$ for which there is only one local
Hamiltonian $\vec{c}$ (up to an overall scalar) that satisfies $K\vec{c}=0$. After many measurements, we can use the
central limit theorem to model the noise as a Gaussian matrix:
\begin{equation}
  \hat{K} - K \approx \epsilon \hat{E},
\end{equation}
where each entry $\hat{E}_{nm}$ of $\hat{E}$ is an independent random variable
with zero mean and unit standard deviation. It is scaled by a small
parameter $\epsilon$ which decays as $n_s^{-1/2}$. 

We wish to estimate the distance between the true and recovered
Hamiltonians. This distance is given by
$\norm{\ket{c_0'}-\ket{c_0}}$, where $\ket{c_0}$ and $\ket{c_0'}$
are the ground-states of the clean $K^TK$ and its noisy estimate
$(K+ \epsilon \hat{E})^T(K + \epsilon\hat{E})$. We treat $K^T K$ as
an unperturbed Hamiltonian, and $\epsilon(\hat{E}^T K + K^T \hat{E})$ as a perturbation to first order in $\epsilon$. We obtain:

\begin{align}
  \ket{c_0'}-\ket{c_0} 
    &= \epsilon\sum_{i>0}\ket{c_i}
      \frac{\broket{c_i}{\hat{E}^TK+K^T \hat{E}}{c_0}}
          {\lambda_{i}-\lambda_{0}}
    + \orderof{\epsilon^2} \\
  &= \epsilon\sum_{i>0}\ket{c_i}
    \frac{\broket{c_i}{K^T \hat{E}}{c_0}}{\lambda_{i}}
      +\orderof{\epsilon^2} ,
\end{align}
where $\ket{c_i}$ are the eigenstates of $K^T K$ and $\lambda_{i}$ the
corresponding eigenvalues in increasing order. $\lambda_0 = 0$ since
we assumed that an exact reconstruction exists, which also implies
that $K$ annihilates $\ket{c_0}$:

\begin{equation}
  K^T K \ket{c_0} = 0 \Rightarrow \norm{K \ket{c_0}}^2 
    = \broket{c_0}{K^T K}{c_0} = 0. 
\end{equation}
Similarly, $\norm{K \ket{c_i}}^2 = \lambda_{i}$, so we can write:
\begin{equation}
K\ket{c_i}
    = \sqrt{\lambda_i} \ket{\tilde{c}_i}
\label{singvec}
\end{equation}
For some unit vector $\ket{\tilde{c}_i}$. Using this we obtain:
\begin{equation}
  \ket{c_0'}-\ket{c_0}
    =\epsilon\sum_{i>0}\frac{1}{\sqrt{\lambda_i}}
      \ket{c_i}\broket{\tilde{c}_i}{\hat{E}}{c_0}
        + \orderof{\epsilon^2},
\end{equation}
and therefore,
\begin{equation}
  \norm{\ket{c_0'}-\ket{c_0}}
    =\epsilon\sqrt{\sum_{i>0}\frac{1}{\lambda_{i}}
      |\broket{\tilde{c}_i}{\hat{E}}{c_0}|^2}
        + \orderof{\epsilon^2}.
\end{equation}
We can now
average over the noise $\hat{E}$ by invoking Jensen's inequality, together
with the concavity of the square root function:
\begin{equation}
\begin{aligned}
  \mathbb{E} \norm{\ket{c_0'}-\ket{c_0}} 
   &= \epsilon \mathbb{E} 
     \sqrt{\sum_{i>0}\frac{1}{\lambda_{i}} 
      |\broket{\tilde{c}_i}{\hat{E}}{c_0}|^2}
    + \orderof{\epsilon^2} \\
  &\le  \epsilon\sqrt{\sum_{i>0}\frac{1}{\lambda_{i}} 
    \mathbb{E}|\broket{\tilde{c}_i}{\hat{E}}{c_0}|^2}
    + \orderof{\epsilon^2} \\
  &= \epsilon\sqrt{\sum_{i>0}\frac{1}{\lambda_{i}}} 
    + \orderof{\epsilon^2} .
\end{aligned}
\end{equation}
In the last equality we used the identity
$\mathbb{E}|\broket{\tilde{c}_i}{\hat{E}}{c_0}|^2=1$, which follows
from:
\begin{equation}
\begin{aligned}
  \mathbb{E}|\broket{\tilde{c}_i}{\hat{E}}{c_0}|^2
  &= \mathbb{E}\left|\sum_{n=1}^N\sum_{m=1}^M
    \tilde{c}_i^nc_0^m\hat{E}_{nm}\right|^2\\
  &= \sum_{n,\nu=1}^N\sum_{m,\mu=1}^M
    \tilde{c}_i^n\tilde{c}_i^{\nu}c_0^m
      c_0^{\mu}\mathbb{E}\big(\hat{E}_{nm}\hat{E}_{\nu\mu}\big)\\
  &= \sum_{n,\nu=1}^N\sum_{m,\mu=1}^M
    \tilde{c}_i^n\tilde{c}_i^{\nu}
      c_0^mc_0^{\mu}\delta_{n,\nu}
        \delta_{m,\mu} \\
  &= \sum_{a=1}^N\sum_{b=1}^M\left|\tilde{c}_i^{a}\right|^2
    \left|c_0^{b}\right|^2 \\\\
  &= \braket{\tilde{c}_i}{\tilde{c}_i}\braket{c_0}{c_0}=1.
\end{aligned}
\end{equation}

\section{Scaling of $\lambda_1$ with sub-system size}
\label{app_XY}

\begin{figure}
\includegraphics[width=8.6cm]{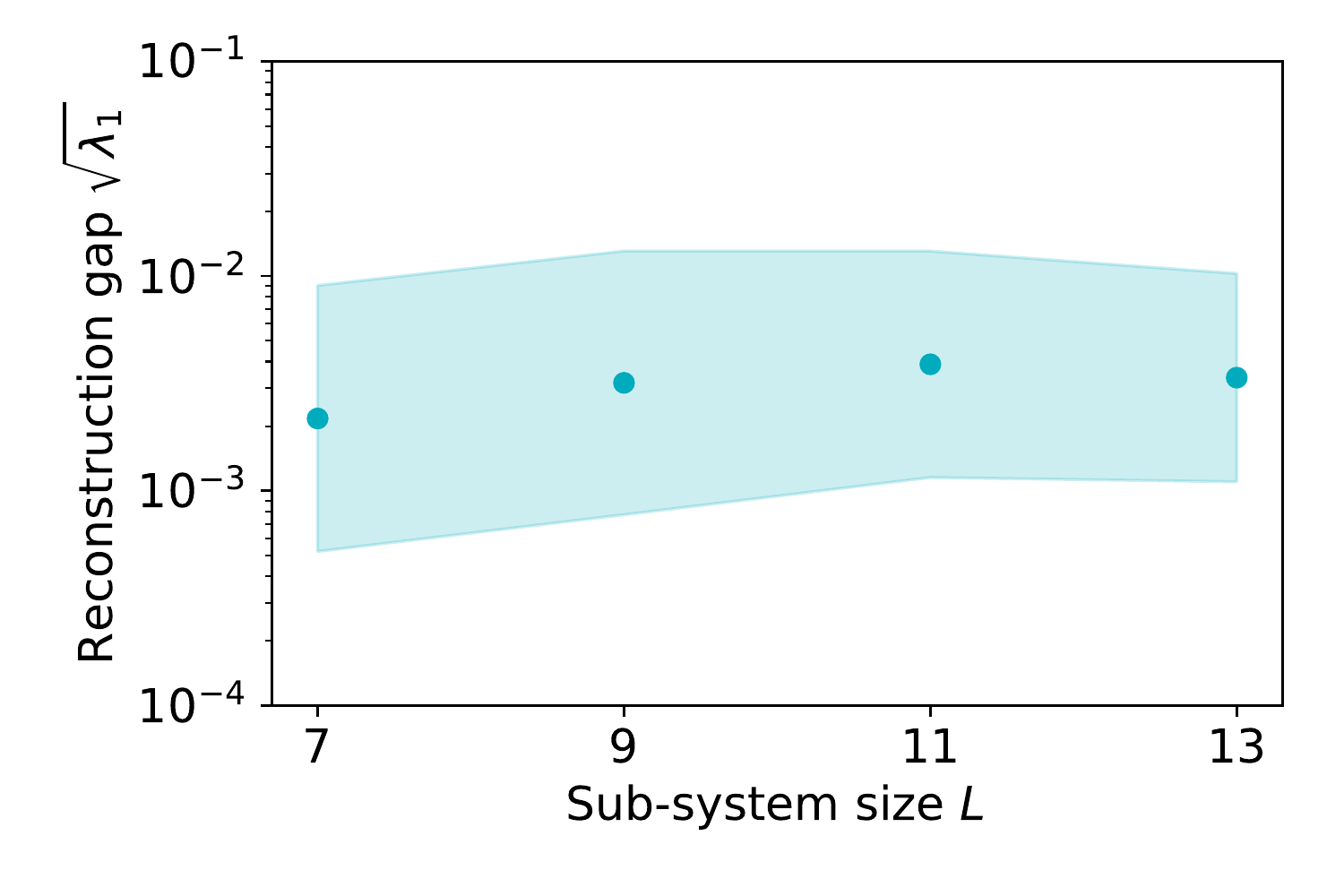}
\protect\caption{Recovering $H_L$ for various sub-system sizes $|L|$ of long $XY$ chains with $|\Lambda=100|$ sites. The reconstruction gap $\lambda_1$, which quantifies the robustness of the reconstruction procedure to noise, seems insensitive to sub-system size for $7 \leq |L| \leq 13$. Results are averaged over 50 randomizations, with mean (dark circles) and standard deviation (light shading) were calculated after taking the log.}
\label{figS1}
\end{figure}

We simulated larger systems of an integrable model to find how the
required number of measurements of each observable scales with
$|L|$. We examined random $XY$ chains, for which the Hamiltonian is given by:
\begin{equation}
  H_{XY} = \frac{1}{2} \sum_{l=1}^{\Lambda} \left[
  2g_l \sigma_{l}^{z} + 
  (1 + \gamma_l) \sigma_{l}^{x} \sigma_{l+1}^{x} + 
  (1 - \gamma_l) \sigma_{l}^{y} \sigma_{l+1}^{y}\right].
\label{Eq:H_XY}
\end{equation}

Using the methods described in \cite{Peschel2003, Cheong2004a,
Peschel2009, Sachdev2011}, we constructed reduced density matrices for
ground-states on $\Lambda=100$ sites. We considered sub-regions
consisting of $|L|=7,9,11,13$ sites, and calculated the gaps of the
correlation matrices $\mathcal{M}$ constructed with all the
$4$-local constraints $A_n$ supported on the corresponding interior
regions $L_0$. The gap of the correlation matrix seems 
insensitive to sub-system size for the sizes we examined (Fig.
\ref{figS1}). 

\section{Recovery for different driving parameters}
\label{app_drive}

\begin{figure*}[t]
\includegraphics[width=18cm]{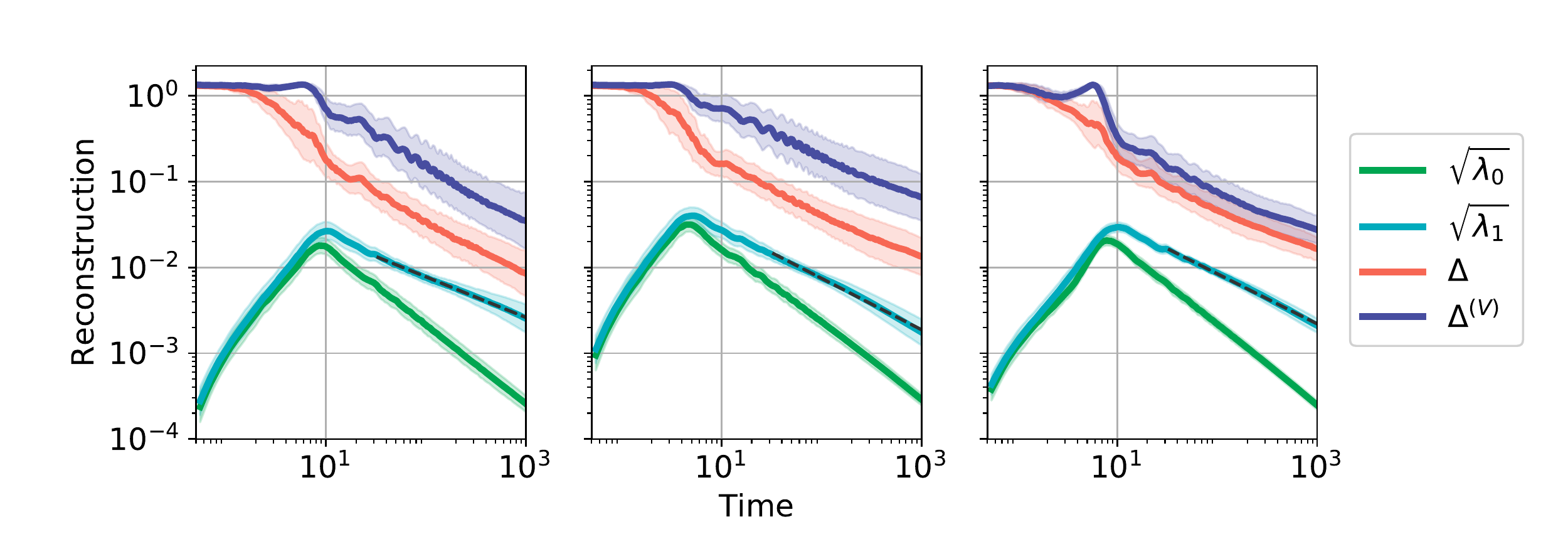}
\protect\caption{Recovery of a time-dependent Hamiltonian as a function of driving amplitude and frequency. Left: right panel of Fig. 4 in the main text (driving frequency $\omega=0.05$, amplitude $J=0.5$). Middle: double frequency ($\omega=0.1$, $J=0.5$). Right: double amplitude ($\omega=0.05$, $J=1$). The power $\alpha$ of the long-time decay of the first excited right-singular value $\sqrt{\lambda_1} \sim t^{-\alpha}$ is larger in the central and right panels compared to the left panel ($-0.599 \pm 0.004$ and $-0.584 \pm 0.003$ compared to $-0.473 \pm 0.002$), indicating more noisy recovery. Indeed, we expect faster heating to arise from a stronger driving amplitude, as well as from a higher driving frequency within this low-frequency regime.}
\label{figS2}
\end{figure*}

Recovery of a time-dependent $H(t)$ is only possible when heating is sufficiently slow. As time progresses, the
time-averaged commutator $[\rho(t), H(t)]$ decays, as quantified by
$\lambda_0$, the lowest right-singular value of the extended
constraint matrix $K_{N \times 2M}$. However, since energy is not conserved for such a system, $\rho _{avg}$ could heat up to an infinite-temperature
fully-mixed state, which trivially commutes with any Hamiltonian.
One measure for this process is the next singular value $\lambda_1$,
quantifying how well any competing Hamiltonian would commute with
$\rho$ on average. Recovery is therefore possible whenever
$\lambda_0$ decays faster than $\lambda_1$, such that the solution
does not mix the true Hamiltonian much with any competitor.

Indeed, we find that stronger or slightly faster driving leads to a more rapid
decay of the reconstruction gap (Fig. \ref{figS2}). This agrees with our expectation that a larger driving
amplitude should lead to faster energy absorption from the drive;
we expect the same from a slightly higher driving frequency within the low-frequency regime we study.

\section{Relation to previously-defined correlation matrix}
\label{app_M}

If we wish to recover the Hamiltonian on the full system $\Lambda$
by enforcing stationarity of all possible constraints $A_n$, the
correlation matrix $\mathcal{M}$ takes the following form, as defined in \cite{Qi2017} (up to a multiplicative scalar):

\begin{equation}
\begin{aligned}
  M_{ij}= & \sum_nK_{ni}K_{nj}\\
    = & \sum_n\left\langle i\left[h_{i},A_n\right]
      \right\rangle \left\langle i\left[h_{j},A_n\right]\right\rangle \\
    = & \sum_n\overline{\Tr\left(\rho\left[h_{i},A_n\right]\right)}\Tr\left(\rho\left[h_{j},A_n\right]\right)\\
    = & \sum_n\overline{\Tr\left(A_n\left[h_{i},\rho\right]\right)}\Tr\left(A_n\left[h_{j},\rho\right]\right)\\
    = & 2^{\Lambda} \Tr\left(\left[h_{i},\rho\right]^{\dagger}\left[h_{j},\rho\right]\right)
\end{aligned}
\end{equation}

Where we used the identity $\Tr \left( A \left[ B, C \right] \right)
= \Tr \left( C \left[ A, B \right] \right) $ which follows from the
cyclic property of the trace, as well as $\left[ A, B \right] = -
\left[ B, A \right] $. Finally, the last equality follows from the
generalized Parseval identity for the Hilbert-Schmidt inner product,
namely:
\begin{equation}
  \left\langle v, w \right\rangle = 
  \sum_{i} \overline{ \left\langle v, e_i \right\rangle }
  \left\langle w, e_i \right\rangle,
\end{equation}
taking the operators $A_n$ to be an orthogonal
basis for the operators on $\Lambda$. We normalize $A_n$ to unity in
operator norm (largest eigenvalue) rather than Hilbert-Schmidt norm,
since measurements in the lab yield $\pm 1$ outcomes; this is the
origin of the $2^{\Lambda}$ factor.

\end{document}